\documentclass[aps,preprint,prd,epsfig]{revtex4}
\usepackage{amssymb}
\usepackage{amsmath}
\usepackage{graphicx}
\usepackage{fancyhdr}
\newcommand{\be}{\begin{equation}}
\newcommand{\ee}{\end{equation}}
\newcommand{\bea}{\begin{eqnarray}}
\newcommand{\eea}{\end{eqnarray}}
\newcommand{\bes}{\begin{subequations}}
\newcommand{\ees}{\end{subequations}}

\newcommand{\bc}{\begin{center}}
\newcommand{\ec}{\end{center}}

\begin{document}

\title{ Realizing the supersymmetric inverse seesaw model in the framework of R-parity violation. }

\author{ C. A. de S. Pires, J. G. Rodrigues,  P. S. Rodrigues da Silva}
\affiliation{{ Departamento de F\'{\i}sica, Universidade Federal da Para\'\i ba, Caixa Postal 5008, 58051-970,
Jo\~ao Pessoa, PB, Brazil}}

\date{\today}

\begin{abstract}
If, on one hand,  the inverse seesaw is the paradigm of TeV scale seesaw mechanism, on the other it is a challenge to find scenarios capable of  realizing it. In this work we propose a scenario, based on the framework of R-parity violation, that realizes minimally  the supersymmetric  inverse seesaw mechanism. In it the energy scale parameters  involved in the mechanism are recognized as the vacuum expectation values  of the scalars that compose the singlet  superfields  $\hat N^C$ and $\hat S$. We develop also the scalar sector of the model and show that  the Higgs mass receives a new  tree-level contribution that, when combined with the standard contribution plus loop correction, is  capable of attaining  $125$GeV without resort to heavy stops.

%
%
\end{abstract}

\maketitle

\section{Introduction}
A current exciting  challenge in particle physics is the explanation of  the smallness of the neutrino masses  through new physics at TeV scale. In this regard, the inverse seesaw mechanism(ISS) \cite{ISS} became the paradigm of successful TeV scale seesaw mechanism. Its minimal implementation requires the introduction to the electroweak  standard model (SM) of two sets of three  neutral fermion singlets , $N=(N_1\,,\,N_2\,,\,N_3) $ and $S=(S_1\,,\,S_2\,,\,S_3)$,  composing the following mass terms in the flavor basis, 
\begin{equation}
{\cal L}_{mass} \supset \bar \nu M_D  N + \bar N M_N S + \frac{1}{2} \bar S^C \mu_N S + h.c,
\label{masstermsoriginal}
\end{equation}
where $\nu=(\nu_1\,,\,\nu_2\,,\,\nu_3)$ is the set of standard neutrinos. In the basis $(\nu\,,\,N^C\,,\,S)$ the neutrino mass  may be put in the following $9 \times 9$  matrix form, 
\begin{equation}
M_\nu=
\begin{pmatrix}
0 & M_D & 0 \\
M^T_D & 0 & M_N\\
0 & M^T_N & \mu_N
\end{pmatrix}.
\label{ISSmatrix}
\end{equation}
In the regime  $\mu_N << M_D < M_N$, the mechanism provides $m_\nu = M_D^T M_N^{-1}\mu_N (M_N^T)^{-1} M_D$ for the mass matrix of the standard neutrinos. Taking $M_D$ at electroweak scale, $M_N$ at TeV and $\mu_N$ at keV scale, the mechanism provides standard neutrinos at eV scale. The new set of fermion singlets $(N\,,\,S)$ develop mass at $M_N$ scale and may be probed at the LHC.

The challenge concerning the ISS mechanism is to find scenarios that  realize it.  This means to propose  models that generate the mass terms in Eq. (\ref{masstermsoriginal}).  In this regard, as the ISS mechanism works in the TeV scale, it seems to be natural to look for realization of the ISS mechanism  in the framework of theories that we expect will manifest at TeV scale\cite{ISSnonsusy1, ISSnonsusy2}, which is the case of supersymmetry ( SUSY). Thus it seems to be interesting to look for  scenarios that realize   the ISS mechanism in the context of SUSY\cite{ISSSUSY1,ISSSUSY2,ISSSUSYR}. 

We know already that a natural way of obtaining small neutrino mass in the context of the MSSM is to consider  that R-parity, $R \equiv (-1)^{2S+3(B-L)}$, is violated  through bilinear terms  like $\mu_i \hat L_i \hat H_u$ in the superpotential\cite{standardRPV}.  Thus we wonder  if  R-parity violation (RPV) is an interesting framework for the  realization of the SUSYISS mechanism.  For this, we implement the SUSYISS mechanism in a framework  where R-parity and lepton number are violated  explicitly but baryon number is conserved in a way that we call the minimal realization of the SUSYISS mechanism once  the necessary  set of superfields required to realize it  is the  original one, $\hat N^C_i$ and $ \hat S_i$, only. 

Moreover, it has been extensively discussed  that the minimal supersymmetric standard model (MSSM) faces difficulties in accommodating a Higgs of mass of 125 GeV, as discovered by ATLAS and CMS\cite{atlasCMS} while keeping the principle of naturalness\cite{higgsmassanalysis}. This is so because, at tree level, the MSSM predicts a Higgs with a mass whose value cannot exceed  $91$ GeV. Thus robust loop corrections are necessary in order to lift this value to $125$ GeV. Consequently stops with mass far above $1$TeV are required. To accept this is to put the naturalness principle aside. We show that the SUSYISS mechanism developed here  accommodates a $125$ GeV Higgs mass  without resort to robust loop corrections.

\section{The mechanism}

The supersymmetric version of the ISS (SUSYISS) mechanism\cite{ISSSUSY1} requires the assumption of two sets of three singlet  superfields $\hat N^C_i\,,\, \hat S_i$ ($i=1,2,3$) composing, with the  MSSM superfields, $\hat L^T_ i =(\hat \nu_i\,,\,\hat e_i)^T\,,\, \hat H_d^T=(\hat H^-_d \,,\,\hat H^0_d )^T\,,\, \hat H_u^T=(\hat H^+_u \,,\,\hat H^0_u)^T $,  the following  extra terms in the superpotential, $W \supset \hat L \hat H_u \hat N^C + \hat S M_N \hat N^C + \frac{1}{2} \hat S \mu_N \hat S$. A successful  extension of the MSSM  that realizes the  SUSYISS mechanism  must generate these terms. This would be an interesting result in particle physics since we would be providing an origin for the energy scales $M_N$ and $\mu_N$\cite{ISSSUSY2}.

The mechanism we propose here is minimal in the sense that it  requires the  addition to the MSSM of the two canonical  singlet superfields $\hat N^C_i $ and $ \hat S_i$, only. Moreover, we  impose that the superpotential be invariant under the following set of discrete symmetries, $Z_3 \otimes Z_2$, according to the following transformation: under $Z_3$ the transformations are,
\begin{equation}
(\hat S_i \,,\, \hat N^C_i\,,\, \hat e^C_i)\,\, \rightarrow \,\,w(\hat S_i \,,\, N^C_i\,,\, \hat e^C_i), \,\,\,\,\hat L_i \,\,\rightarrow \,\, w^2 \hat L_i,
\label{z3}
\end{equation}
with $w=\exp^{i2\pi/3}$. Under   $Z_2$ we have,   $\hat S_i \rightarrow - \hat S_i$, with all the remaining superfields transforming trivially by $Z_3\otimes Z_2$. 

Thus the superpotential of the  SUSYISS mechanism we propose here involves the following  terms,
\small
\begin{eqnarray}
\hat{W}&=&\mu \hat{H}^{a}_{u}\hat{H}_{da} +
Y^{ij}_{u}\epsilon_{ab}\hat{Q}^{a}_{i}\hat{H}^{b}_{u}\hat{u}^{c}_{j} +
Y^{ij}_{d}\hat{Q}^{a}_{i}\hat{H}^{b}_{d}\hat{d}^{c}_{j} +
Y^{ij}_{e}\hat{L}^{a}_{i}\hat{H}^{b}_{d}\hat{e}^{c}_{j} \nonumber \\ 
&+&
Y^{ij}_{\nu}\epsilon_{ab}\hat{L}^{a}_i\hat{H}^{b}_{u}\hat{N}^{c}_{i} + \frac{1}{2}\lambda^{ijk}_{s}\hat{N}^{c}_{i}\hat{S}_{j}\hat{S}_{k} + \frac{1}{3}\lambda^{ijk}_{v}\hat{N}^{c}_{i}\hat{N}^{c}_{j}\hat{N}^{c}_{k},
\label{superpotential}
\end{eqnarray}
where $a \,,\,b$ are $SU(2)$ indices and $i$ and $j$ are generation indices.  $\hat{Q}_{i}$,  $\hat{u}^{c}_i$, $\hat{d}^{c}_{i}$ and $\hat{e}^{c}_{i}$ are the standard superfields of the MSSM. Perceive that the $Z_3 \otimes Z_2$ symmetry permits that lepton number as well as R-parity be explicitly violated in this model by  terms in the superpotential that involve the singlet superfields  $\hat N^C_i $ and $ \hat S_i$, only .

Now we make an important assumption. We assume  that the  scalars that compose the   superfields  $\hat N^C_i $ and $\hat S_i$ develop nonzero vacuum expectation value (VEV),  $\langle \tilde S \rangle = v_{S_i}$ and $ \langle \tilde N^C_i \rangle =v_{N_i}$, respectively. This assumption provides the source of the canonical mass terms $M_N$ and $\mu_N$ of the SUSYISS mechanism. Note that,   from the last  two  terms in the superpotential above, we have that   the VEV of the scalar $ \tilde S$ becomes the source of the mass scale $M_N$ while the VEV of the scalar $ \tilde N^C$ becomes the source of the mass scale $\mu_N$ . In other words, the superpotential above together with the assumption that the scalars $\hat N^C_i $ and $\hat S_i $ develop non zero  VEVs has the required ingredients  to realize the SUSYISS mechanism.

Another important point of the model is to discuss  the possible values  $v_{S_i}$ and $v_{N_i}$ may take. For this we have to obtain the potential of the model.  The soft breaking sector will play an important role in the form of the potential.

The most general soft breaking sector of our interest involves the following terms,
\begin{eqnarray}
-\cal{ L}_{\mbox{soft}} &=& M^{2}_{Q_{ij}}\tilde{Q^a_i}^{*}\tilde{Q^a_j}
+ M^{2}_{u^{c}_{ij}}\tilde{u^c_i}^{*}\tilde{u^c_j} +
M^{2}_{d^{c}_{ij}}\tilde{d^c_i}^{*}\tilde{d^c_j} \nonumber \\
&+&
M^{2}_{L_{ij}}\tilde{L^a_i}^{*}\tilde{L^a_j} +
M^{2}_{e^{c}_{ij}}\tilde{e^c_i}^{*}\tilde{e^c_j} +
M^{2}_{h_{u}}H^{a*}_u H^{a}_u \nonumber \\
&+& M^{2}_{h_{d}}H^{a*}_d H^{a}_d + M^{2}_{\tilde N_i}
\tilde{N_i}^{*C}
\tilde{N}^{C}_i + M^{2}_{\tilde S_i}
\tilde{S^*_i} \tilde{S_i}
\nonumber \\
&-&
[\left(A_{u}Y_{u}\right)_{ij}\epsilon_{ab}\tilde{Q}^{a}_{i}H^{b}_{u}\tilde{u}^{c}_{j}
+ \left(A_{d}Y_{d}\right)_{ij}\tilde{Q}^{a}_{i}H^{a}_{d}\tilde{d}^{c}_{j}
\nonumber \\
&+&
\left(A_{e}Y_{e}\right)_{ij}\tilde{L}^{a}_{i}H^{a}_{d}\tilde{e}^{c}_{j} +
h.c.] - [B\mu H^{a}_u H^{a}_d + h.c.] \nonumber \\
&+& \frac{1}{2}
\left(M_{3}\tilde{\lambda}_{3}\tilde{\lambda}_{3}
+ M_{2}\tilde{\lambda}_{2}\tilde{\lambda}_{2} +
M_{1}\tilde{\lambda}_{1}\tilde{\lambda}_{1} + h.c.\right) \nonumber \\ 
&+& (A_{y}Y_{\nu})^{ij}
\epsilon_{ab}\tilde{L}^{a}_i  H^{b}_{u}\tilde{N}^{*C}_j \nonumber \\
&+& [\frac{1}{2}
(A_{s}\lambda_{s})^{ijk}\tilde{N}^{*C}_i\tilde{S_j}\tilde{S_k} +
\frac{1}{3}(A_{v}\lambda_{v})^{ijk}\tilde{N}^{* C}_i\tilde{N}^{*C}_j
\tilde{N}^{*C}_k  \nonumber \\ &+& h.c.].
\label{softterms}
\end{eqnarray}
Note that the last two trilinear terms violate explicitly lepton number and the energy scale parameters $A_s$ and $A_v$ characterize such violation.

A common assumption in developing ISS mechanisms it  to assume that the new neutral singlet fermions are degenerated in masses and self-couplings. However, for our case here, it seems to be more convenient,  instead of considering the degenerated case, to consider the case of only one generation of superfields. The extension for the case of three generations is straightforward and the results are practically the same. 

The potential of the model is  composed by the terms $V=V_{soft} + V_D + V_F$. The soft
term, $V_F$,  is given  above in Eq. (\ref{softterms}). The relevant contributions to $V_D$  are,
\begin{equation}
V_{D}= \frac{1}{8}(g^{2}+g^{\prime 2})(\tilde{\nu}\tilde{\nu}^* + H^0_d H^{0*}_d -
H^0_u H^{0*}_u)^2.
\label{Dterm}
\end{equation}

In what concerns the  F-term, the relevant contributions are given by the following terms,

\begin{eqnarray}
V_{F} &=& \left|\frac{\partial \hat{W}}{\partial \hat{H^{0}_{u}}
}\right|^{2}_{H_{u}} + \left|\frac{\partial \hat{W}}{\partial
\hat{H^{0}_{d}} }\right|^{2}_{H_{d}} + \left|\frac{\partial \hat{W}}{\partial
\hat{\nu}} \right|^{2}_{\tilde{\nu}} + \left|\frac{\partial \hat{W}}{\partial
\hat{N^{C}} }\right|^{2}_{\tilde{N}} + \left|\frac{\partial
\hat{W}}{\partial \hat{S}_{L}}\right|^{2}_{\tilde{S}} \nonumber \\
&=& \mu^{2}\left|H^{0}_{u}\right|^{2} + \mu^{2}\left|H^{0}_{d}\right|^{2} +
Y^{2}_{\nu}\left|\tilde{N}^{C}\right|^{2}\left|\tilde{\nu}\right|^{2} + Y_{v}\mu
H^{0*}_{d}\tilde{N}^{C*}\tilde{\nu}
\nonumber \\
&+& Y^{2}_{\nu}\left|H^{0}_{u}\right|^{2}\left|\tilde{\nu}\right|^{2}
+ \frac{1}{4}\lambda^{2}_{s}\left|\tilde{S}\right|^{4} + 4
\lambda^{2}_{v}\left|\tilde{N}^{C}\right|^{4} +
\lambda^{2}_{s}\left|\tilde{N}^{C}\right|^{2}\left|\tilde{S}\right|^{2}\nonumber \\
&+& 
\frac{Y_{\nu}\lambda_{s}H^{0}_{u}\tilde{\nu}\tilde{S}^{* 2}}{2} +
2Y_{\nu}\lambda_{v}\left|\tilde{N}^{C}\right|^{2}H^{0}_{u}\tilde{\nu} +
Y^{2}_{\nu}\left|\tilde{N}^{C}\right|^{2}\left|H^{0}_{u}\right|^{2} \nonumber \\ 
&+&
\lambda_{s}\lambda_{v}\left|\tilde{N}^{C}\right|^{2}\tilde{S}^{2} + h.c.
\label{Fterm}
\end{eqnarray}

With the potential of the model in hand,  we are ready  to obtain the  set of constraint equations for the neutral scalars $H^0_u\,,\, H^0_d\,,\,\tilde \nu\,,\, \tilde S\,,\,\tilde N^C$,
\begin{eqnarray}
&&  v_u\left( M^{2}_{h_u} + \mu^{2} +
\frac{1}{4}(g^{2}+g^{\prime 2})(v^{2}_{u}-v^{2}_{d}-v^2_\nu)  +Y^2_\nu v^2_N +Y^2_\nu v^2_\nu \right) +\nonumber \\
&& -B\mu v_d+ \frac{1}{2}Y_{\nu}\lambda_{s}v_\nu v^{2}_{S}+Y_\nu A_y v_\nu v_N  + 2Y_\nu\lambda_v v_\nu v^2_N =0,\nonumber
\\
&& v_d\left(M^{2}_{h_d} + \mu^{2} -
\frac{1}{4}(g^{2}+g^{\prime 2})(v^{2}_{u}-v^{2}_{d}-v^2_\nu) \right) -B\mu v_u+
Y_\nu \mu v_\nu v_N=0,\nonumber
\\
&&  v_\nu \left(M^{2}_{\tilde \nu} +
\frac{1}{4}(g^{2}+g^{\prime 2})(v^2_\nu + v^{2}_{d}-v^{2}_{u})+Y^{2}_{\nu}v^{2}_{u} +
Y^{2}_{\nu}v^{2}_{N}\right) + \nonumber \\
&& + \frac{1}{2}\lambda_{s}Y_{\nu}v_{u}v^{2}_{S} + Y_\nu A_y v_u v_N + 2Y_\nu
\lambda_v v_u v^2_N + Y_\nu \mu v_d v_N =0, \nonumber \\
&& M^{2}_{\tilde S} + \lambda_{s}Y_{\nu}v_{u}v_\nu + \frac{1}{2}
\lambda^{2}_{s} v^{2}_{S} +\lambda_s A_s v_N + \lambda^2_s v^2_N  +
2\lambda_s\lambda_v v^2_N=0,\nonumber
\\
&&v_{N}\left( M^{2}_{\tilde N} + Y^{2}_{\nu}v^{2}_{u} + \lambda^{2}_{s}v^{2}_{S} +
3\lambda_v A_v v_N + 8\lambda^2_v v^2_N + 4\lambda_{v}Y_{\nu} v_{u} v_\nu
+ 2\lambda_{v}\lambda_{s}
v^{2}_{S} + Y^{2}_{\nu}v^2_\nu \right) +\nonumber \\
&&+ Y_\nu v_\nu (A_{y}v_{u}   + \mu v_{d})+\frac{1}{2} A_{s} \lambda_{s} v^{2}_{S}=0.
\label{constraints}
\end{eqnarray}

Let us first focus on the third relation in the equation above. Observe that the dominant term inside the parenthesis is $M^{2}_{\tilde \nu} $. Outside the parenthesis, on considering for while that $v_N < v_S$, the dominant term is $\frac{1}{2}\lambda_{s}Y_{\nu}v_{u}v^{2}_{S}  $. In view of this, from the third relation above  we have that,
\begin{equation}
v_\nu \approx -\frac{1}{2}
\frac{\lambda_{s}Y_{\nu}v_{u}v^{2}_{S}}{M^2_{\tilde \nu}}.
\label{vnu}
\end{equation}
 For  $M_{\tilde \nu} > v_S$, we have $v_\nu < v_{u , d, S} $, as expected.

Let us now focus on  the fifth relation of the Eq. (\ref{constraints}). The dominant term inside the parenthesis is $M^{2}_{\tilde N}$, while  outside the parenthesis the dominant term  is $\frac{1}{2} A_{s} \lambda_{s} v^{2}_{S}$. Thus the fifth relation  provides,
\begin{equation}
v_N \approx -\frac{1}{2}\frac{ A_{s} \lambda_{s} v^{2}_{S}}{M^2_{\tilde N}} .
\label{smallseesaw}
\end{equation}
 This expression for $v_N$ is similar to the $v_\nu $ case and suggests that $v_N$ is also small. 
 
 Let us now focus on the forth relation. Taking $v_\nu\,,\,v_N \ll v_S$, we have that the dominant terms in that relation are,
 \begin{equation}
 M^{2}_{\tilde S} + \frac{1}{2}
\lambda^{2}_{s} v^{2}_{S} =0.
\label{vS}
\end{equation}
Perceive that  $M_S$ dictates the value of $v_S$. As the neutral singlet scalar $\tilde S$ belongs to an extension of the MSSM, then it is reasonable to expect that  its soft mass term $M_S$ lies at TeV scale. Consequently  $v_S$  must assume values around TeV. In regard to the first and second relations they control the standard VEVs $v_u$ and $v_d$.

Let us return to the expression to $v_N$ in Eq. (\ref{smallseesaw}). As the neutral singlet scalar $\tilde N$ also belongs to an extension of the MSSM, then it is reasonable to expect that  its soft mass term $M_{\tilde N}$  lies at TeV scale, too. In this case perceive that the value of $v_N$ get dictated by the soft trilinear term $A_s$. Thus a small $v_N$ means a tiny $A_s$. As $A_s$ is a trilinear soft breaking term, then it must be generated by some spontaneous SUSY breaking scheme. The problem is that we do not know how SUSY is spontaneously  broken. Thus there is no way to infer exactly the value  of $A_s$.   Moreover, note that $A_s$ is a soft trilinear term involving only neutral  scalar singlets by MSSM which turns its estimation even more complex. We argue here that it is somehow natural to expect that such terms be  small.

For this we have to think in terms of spontaneous SUSY breaking schemes.  For example, in the framework of gauge mediated supersymmetry breaking (GMSB) all soft trilinear terms are naturally suppressed once they arise from  loops. In our case  the new singlets are sterile by the standard gauge group.  The  minimal scenario where such soft trilinear terms could arise would be one that involve the GMSM of the B-L gauge extension of the MSSM. To build such extension and evaluate $A_s$ in such a scenario is out of the scope of this paper. However, whatever be the case,  in the framework of GMSB scheme   $A_s$ must be naturally small and consequently  $v_N$, too.  In this point we call the attention to the fact that the idea behind the ISS mechanism is that lepton number is explicitly violated at low energy scale.  This suggests  that the GMSB seems to be the adequate spontaneous SUSY breaking scheme to be adopted in realizing  SUSYISS mechanism.

Let us discuss the case of  gravity mediated supersymmetry breaking.  As in the ISS mechanism  lepton number is assumed to be explicitly violated at low energy scale, it is expected that    $v_N\,,\,v_S\,,\, v_\nu \, ,\, A_s \,,\,A_v$ are all null at GUT scale.  Considering this, the authors of  Ref. \cite{SUSYISSvalle} evaluated  the running of soft trilinear terms involving scalar singlets  from GUT to down scales in a different realization of the SUSYISS model. As a result they obtained that these terms  develop small values at electroweak scale.   Our case is somehow similar to the case of Ref.  \cite{SUSYISSvalle} and it seems reasonable to expect that, in the general case of three generations, on doing such  evaluation of the running of the soft trilinear terms, our mechanism recover the  result of Ref.  \cite{SUSYISSvalle}.  As we are just presenting the idea by means  of only one generation, such evaluation  of the running of $A_s$ is out of the scope of this work.

Thus it seems to be reasonable to expect that, whatever be the spontaneous SUSY breaking scheme adopted, the soft trilinear terms that violate explicitly lepton number involving  neutral singlet fields as  $\tilde S$ and $\tilde N$ have the tendency to develop  small values.  In what follow we assume  that $A_s$ and $A_v$  lies at keV scale.
 
There is still an issue to consider in respect to the scalar potential. As
can be easily verified, the value of the potential at origin of the fields is zero. 
In order to guarantee that electroweak symmetry will be broken,  we need the potential in the  minimum  to be negative. Taking the constraints in Eq. (\ref{constraints}) to eliminate the soft masses in the
scalar potential, we have,
\begin{eqnarray}
\langle V \rangle_{mim} = &-&\frac{1}{8}\left(g^{2}+g'^{2}\right)\left(v^{2}_{\nu} + v^{2}_{d} -
v^{2}_{u}\right)^2 - Y^{2}_{\nu}\left( v^{2}_{\nu}v^{2}_{N} +
v^{2}_{\nu}v^{2}_{u} + v^{2}_{u}v^{2}_{N}\right) -
\lambda^{2}_{s}v^{2}_{S}v^{2}_{N} - \frac{1}{4}\lambda^{2}_{s}v^{4}_{S} - A_{y}Y_{\nu}v_{\nu}v_{N}v_{u} \nonumber
\\ &-&
\frac{1}{2}A_{s}\lambda_{s}v_{N}v^{2}_{S} - A_{v}\lambda_{v}v^{3}_{N}
-Y_{\nu}\lambda_{s}v_{\nu}v_{u}v^{2}_{S} -
4Y_{\nu}\lambda_{v}v_{\nu}v_{u}v^{2}_{N} -
2\lambda_{s}\lambda_{v}v^{2}_{N}v^{2}_{S} - 4\lambda^{2}_{v}v^{4}_{N} \nonumber
\\ &-& Y_{\nu}\mu v_{\nu}v_{N}v_{d}.
\end{eqnarray}
For the magnitudes of VEVs discussed above, the dominant term is $-
\frac{1}{4}\lambda^{2}_{s}v^{4}_{S}$, which is negative. For the case of one generation considered here this is a strong evidence of the   stability of the potential. 

After all these considerations, we are ready to go to the central part of this work that is  to develop the neutrino sector. For this we have, first, to obtain the mass matrix that involves the neutrinos. Due to the RPV the gauginos and Higgsinos  mix with the neutrinos $\nu$, $N$ and $S$. Considering the basis $(\lambda_{0},\lambda_{3},\psi_{h^{0}_{u}}, \psi_{h^{0}_{d}},\nu,N^{c},S)$, we obtain the following mass matrix for these neutral fermions,
\begin{eqnarray}
 \left(\begin{array}{ccccccc}  M_{1} &
  0 & \frac{g' v_{u}}{\sqrt{2}} & -\frac{g' v_{d}}{\sqrt{2}} & -\frac{g'
  v_\nu}{\sqrt{2}} & 0 & 0 \\
 0 & M_{2} & -\frac{g v_{u}}{\sqrt{2}} & \frac{g v_{d}}{\sqrt{2}} & \frac{g v_\nu}{\sqrt{2}} & 0 & 0\\
 \frac{g' v_{u}}{\sqrt{2}}  & -\frac{g v_{u}}{\sqrt{2}}  &  0 & \mu & Y_{\nu} v_{N} & Y_{\nu} v_\nu & 0 \\
 -\frac{g' v_{d}}{\sqrt{2}}  & \frac{g v_{d}}{\sqrt{2}} & \mu & 0 & 0 & 0 & 0 \\
 -\frac{g'  v_\nu}{\sqrt{2}} & \frac{g v_\nu}{\sqrt{2}} & Y_\nu v_N & 0 & 0 & Y_{\nu} v_{u} & 0 \\
 0 & 0 & Y_\nu v_\nu & 0 & Y_{\nu} v_{u} & 6 \lambda_{v} v_{N} & \lambda_{s} v_{S} \\
 0 & 0 & 0 & 0 & 0 & \lambda_{s} v_{S}  & \lambda_{s} v_{N}  
 \end{array}\right),
 \label{generalneutrinomassmatrix}
\end{eqnarray}
where  $M_{1}$ e $M_{2}$ are the standard  soft breaking terms of the gauginos. We remark that  on considering the hierarchy $v_N< v_\nu < v_d< v_u<v_S$  the bottom right $3 \times 3$ block of this matrix, which   involves only  the neutrinos, decouples from the  gauginos and Higgsinos sector leaving the neutrinos with the following mass matrix in the basis $(\nu,N^{c},S)$
\begin{eqnarray}
M_\nu \approx  \left(\begin{array}{ccc} 
 0 & Y_{\nu} v_{u} & 0 \\
 Y_{\nu} v_{u} & 2 \lambda_{v} v_{N} & \lambda_{s} v_{S} \\
 0 & \lambda_{s} v_{S} & \lambda_{s} v_{N}  
 \end{array}\right).
 \label{neutrinomassmatrix}
\end{eqnarray}

For this decoupling to be effective we must have  
$v_\nu$ of order MeV or less. Diagonalization of this mass matrix implies
that the lightest neutrino, which is predominantly the standard one, $\nu$, get the following mass expression,
\begin{equation}
m_\nu \approx \frac{Y_{\nu}^2}{\lambda_s}\frac{ v_{u}^2}{ v_{S}^2}  v_{N} .
\label{ISSmass}
\end{equation}

This is exactly  the mass expression of the ISS mechanism.  For $v_S$ around 
TeV and $v_N$ around keV we obtain neutrinos at eV scale for  $v_u$ at
electroweak scale. In the case of three generations the pattern of the neutrino
masses will be determined by $Y^{ij}_\nu$ . 

To demonstrate the validity of
these aproximations we can compute the mass eigenvalues from the full matrix in
(\ref{generalneutrinomassmatrix}). For typical values of the supersymmetric parameters and $v_S \sim 10$ TeV, $v_N \sim 10$ keV, $v_\nu \sim 1$ MeV and  $Y_\nu \sim \lambda_s=0.21$, we have the following order of magnitude for the  mass
eigenvalues ($\sim$ TeV, $\sim$ TeV,
$\sim 10^2$ GeV, $\sim 10^2$ GeV, $\sim 10^{-1}$ eV, $\sim$ TeV, 
$\sim$ TeV), where the lightest particle is  exclusively
the standard neutrino. This result is encouraging and indicates that RPV is an interesting framework to realize the SUSYISS mechanism.

We end this section making a comparison of the SUSYISS developed here with the $\mu\nu$SSM in Ref. {\cite{munucase}}. This model resorts to R-parity violation to solve the $\mu$ problem. However neutrino masses at sub-eV scale require considerable amount of fine tuning of the Yukawa couplings. We stress that, in spite of the fact that  the SUSYISS model contains the particle content of the $\mu\nu$SSM, unfortunately it  does not realize the $\mu\nu$SSM.  This is so because if we allow a term like $\hat S \hat H_u \hat H_d$ in the superpotential in Eq. (\ref{superpotential}), as consequence the entries  $\psi_{h^{0}_{d}} S$ and $\psi_{h^{0}_{u}} S$ in the mass matrix in Eq. (\ref{generalneutrinomassmatrix}) would develop robust values which jeopardize the realization of the ISS mechanism. 

\section{The mass of the Higgs}

Now, let us focus on the scalar sector of the model. We restrict our interest   in checking if the model may accommodates   a $125$ GeV Higgs mass  without resorting to  tight loop contributions.  For the case of  one generation the model involves five neutral scalars whose mass terms compose a  $5\times 5 $  mass matrix that we consider in the basis $(H_u\,,\,H_d\,, \tilde \nu \,,\, \tilde N \,,\, \tilde S)$. We do not show such a mass matrix here because of the complexity of  their entries.  Instead of dealing with a $5 \times 5$ mass matrix, which is  very difficult to handle analytically, we make use of a result that says that an upper bound on the mass of the lightest scalar,  which we consider as the Higgs, can be obtained by computing the eigenvalues of the $2 \times 2$ submatrix in the upper left corner of this $5 \times 5$ mass matrix\cite{upperbound}. This is a common procedure adopted in such cases which  give us an idea of the potential of the model to generate the 125 GeV Higgs  mass.  

The dominant terms of this $2 \times 2$ submatrix  are given by,
\begin{eqnarray}
M^{2}_{2\times 2}\approx \left(\begin{array}{cc} B\mu\cot(\beta)+
M^{2}_{Z}\sin^{2}(\beta)-\frac{Y_{\nu}\lambda_{s}v_\nu}{2v_u}
v^2_{S} & -B\mu-M^{2}_{Z}\sin(\beta)\cos(\beta)\\
-B\mu-M^{2}_{Z}\sin(\beta)\cos(\beta) & B\mu \tan(\beta) + M^{2}_{Z}\cos^{2}(\beta)  \end{array}\right).
\label{2x2massmatrix}
\end{eqnarray}
We made use of the hierarchy among the VEVs, as discussed above, to obtain such a $2 \times 2$ submatrix.  On diagonalizing this $2 \times 2$  submatrix we obtain  the following upper bound on the mass of the Higgs,
\begin{equation}
m^{2}_{h}\leq 
M^{2}_{Z}\cos^2(2\beta)-\frac{Y_{\nu}\lambda_{s}v_\nu}{2v_u}v^2_{S}.
\label{upperbound}
\end{equation}
Note also  that  Eq. ({\ref{vS}}) imposes that either  $M^2_{\tilde
S}$ or $v^2_{S}$ is negative. In order to the second term in Eq. (\ref{upperbound})  gives  a positive contribution to the Higgs mass we  take $M^2_S$ negative and $Y_\nu$ and $\lambda_s$ with opposite sign.

What is remarkable in the mass expression in Eq. (\ref{upperbound}) is that the second term provides a robust correction to the Higgs mass even  involving the parameters that dictate the neutrino masses as the
couplings $Y_\nu$ and $\lambda_s$ and the VEV $v_S$. This  suggest an interesting connection between
neutrino and Higgs mass.  For illustrative proposals, perceive that for
$Y_\nu$ of the same order of $\lambda_s$, $v_\nu$ around MeV, $v_u$ around
$10^2$ GeV and $v_{S}$ of order tens of TeV, the second term provides a contribution of tens of
GeV to the Higgs mass. This contribution is enough to alleviate the pressure on
the stop masses and their mixing in order to keep valid the principle of naturalness.

In order to check the range of values the stop mass and the $A_t$ term may
develop in this model, we add to  $m^{2}_{h}$ given above  the leading 1-$\textit{loop}$ corrections
coming from the MSSM stop terms\cite{loopcorrection},
\begin{equation}
\Delta m^{2}_{h}=
\frac{3m^{4}_{t}}{4\pi^{2}v^{2}}\left(log\left(\frac{m^{2}_{s}}{m^{2}_{t}}\right)
+
\frac{X^{2}_{t}}{m^{2}_{s}}\left(1-\frac{X^{2}_{t}}{12m^{2}_{s}}\right)\right),
\label{Limh2}
\end{equation}
where $m_{t}=173.2$ GeV is the top mass, $v=\sqrt{v^2_u + v^2_d}=174$ GeV is the VEV
of the standard model, $X_{t}\equiv A_t-\mu cot (\beta)$ is the stop mixing parameter and $m_{s} \equiv
(m_{\tilde{t}_{1}}m_{\tilde{t}_{2}})^{1/2}$ is the SUSY scale (scale of
superpartners masses) where $m_{\tilde t}$ is the stop mass. In the analysis done below, we work with
degenerated stops and, in all plots, we take $v_\nu=1$ MeV and
$v_{S}=4 \times 10^4$ GeV.

 Figure 1 shows  possible values for the magnitude of $Y_\nu$ and $\lambda_s$
 that provide a Higgs with a mass of $125$ GeV. Note that the plot tells us that
 such a mass requires $Y_\nu$ and $\lambda_s$ around
 $10^{-1}$. This range of values for $Y_\nu$ and $\lambda_s$ provides, through
 Eq.(\ref{ISSmass}),   $m_\nu \approx 0.1$ eV for $v_S=10$ TeV and $v_N=10$ keV.
 Thus neutrino at sub-eV scale is compatible with $m_h=125$ GeV effortlessly.

 Figure 2  tell us  that the model yields the desired Higgs mass for stop mass
 and mixing parameters below the TeV scale. Finally, Figure 3  shows that  a
 Higgs of mass of $125$ GeV is obtained for a broad range of values of $tan (\beta)$.

Let us discuss a little some phenomenological aspects of the SUSYISS mechanism
developed here. First of all, observe that the aspects of RPV concerning the
mixing among neutralinos and neutrinos, as well as charginos and charged
leptons, are dictated by the VEVs $v_\nu$ and $v_{N}$ and the couplings $Y_\nu$
and $\lambda_s$, which are both small. The squarks sector is practically
unaffected. Thus, with relation to these sectors, the phenomenology of the
SUSYISS mechanism is practically similar to the case  of the supersymmetric
version of the ISS mechanism\cite{ISSSUSY1, ISSpheno}. The signature of the
SUSYISS mechanism developed here should manifest mainly in the  scalar sector of
the model due to the mixing of the neutral scalars with the sneutrinos which
will generate Higgs decay channel with lepton flavor violation $h \rightarrow
l_i l_j$.

In general, as far as we know, this is the first time the ISS mechanism is
developed in the framework of RPV. Thus many theoretical, as well
phenomenological aspects of the model proposed here must be addressed such as
experimental constraints from RPV, accelerator physics, analysis of the
renormalization group equation, spontaneously SUSY breaking schemas,  etc.,
which we postpone to a future paper\cite{future}. Moreover, needless to say that
in SUSY models with RPV the lightest supersymmetric particle is not stable which
means that neither the neutralino nor sneutrino are  candidates for dark
matter\cite{DMcandidate} any longer. We would like to remark that
because of the $Z_3$ symmetry used in the superpotential above cosmological
domain wall problems are expected\cite{DWproblem}. However, the solution of this
problem  in the NMSSM as well in  the $\mu\nu$SSM\cite{munucase} cases may be
applied to our case, too\cite{DWsolution}. 

Finally, concerning the stability of the vacuum, we have to impose that the potential be bounded from below when the scalar fields become  large in any direction of the space  fields and that the potential does not present charge and color breaking  minima. Concerning the latter condition, we do not have to worry about this condition here because the new scalar fields associated to the superfield singlets, $\hat S$ and $\hat N^C$,  are neutrals under electric and color charges. Concerning the former issue, the worry arises because at large values of the fields the quartic terms dominate the potential. Thus we have to guarantee  that  at large values of the fields the potential be positive. Thus we have to worry with the quartic couplings, only.  The negative value of $\lambda_s$  leads to two negative quartic  terms. Considering this, on analyzing the potential above, we did not find any direction in the field space in which $\lambda_s$ negative leads to a negative potential. All direction we find involves a set of condition where it is always possible to guarantee that the potential be positive at large value of the fields\cite{casas}.  Moreover, we took $\lambda_s$ negative for convenience. We may arrange the things such that all couplings be positive. For example, on taking $\lambda_s$ positive, $v_\nu$  in Eq. (\ref{vnu}) get negative, which guarantee a positive contribution to the Higgs masses and that all quartic couplings be positive. However, a complete analysis of the stability of the potential is necessary. This  will be done in \cite{future}.

\section{ Conclusions}

In this work we proposed the realization of the SUSYISS model in the framework
of RPV. The main advantage of such framework is that it allows  the realization
of the SUSYISS model with a minimal set of superfield content where the
superfields $\hat S$ and $\hat N^C$ of the minimal implementation are sufficient
to realize the model. To grasp the important features of the SUSYISS,  we
restricted our work to the case of one generation of superfields.   As nice
result, the canonical mass parameters $M_N$ and $\mu_N$ of the SUSYISS mechanism
are recognized as the VEVs of the scalars $\tilde S$ and $\tilde N$ that compose
the superfields $\hat S$ and $\hat N^C$. There is no way to fix the values of
the VEVs $v_S$ and $v_N$. However, it seems plausible that $v_S$ and $v_N$
develop values around TeV and keV scale, respectively. Thus, we conclude that 
RPV seems to be an interesting framework for the realization of the SUSYISS
mechanism. We recognize that in order to establish the model a lot of work have
to be done, yet. For example, we have to  find the spontaneous SUSY breaking
scheme that better accommodates the mechanism, develop the phenomenology of
the model and its embedding in GUT schemes. We end by saying  that the main
results of this work are that  the model proposed here realizes minimally the
SUSYISS mechanism and provides a 125 GeV Higgs mass respecting the naturalness principle.

\acknowledgments
This work was supported by Conselho Nacional de Pesquisa e
Desenvolvimento Cient\'{i}fico- CNPq (C.A.S.P and  P.S.R.S ) and Coordena\c c\~ao de Aperfei\c coamento de Pessoal de N\'{i}vel Superior - CAPES (J.G.R). 

%
\newpage
\begin{figure}[tb]
\begin{center}
	\includegraphics[scale=0.96]{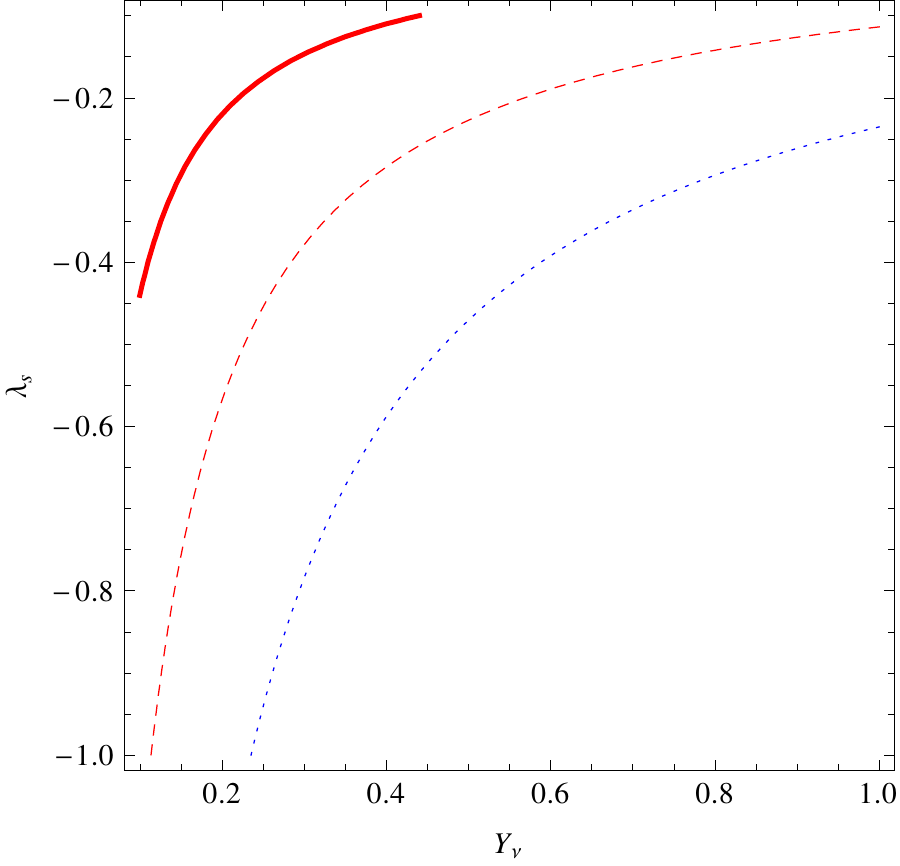} \quad
\caption{Contour plot  of $m_h=125$GeV  in the  $Y_\nu \, \mbox{versus} \,
\lambda_s$ plane for   $m_s=800$GeV and $X_t=400$GeV where (blue dotted  $tan (\beta) =5$), ( red dashed   $tan (\beta)=7$) and (red solid $tan (\beta) =10$).}
 \label{fig1}
\end{center}
\end{figure}
\begin{figure}[tb]
\begin{center}
	\includegraphics[scale=0.96]{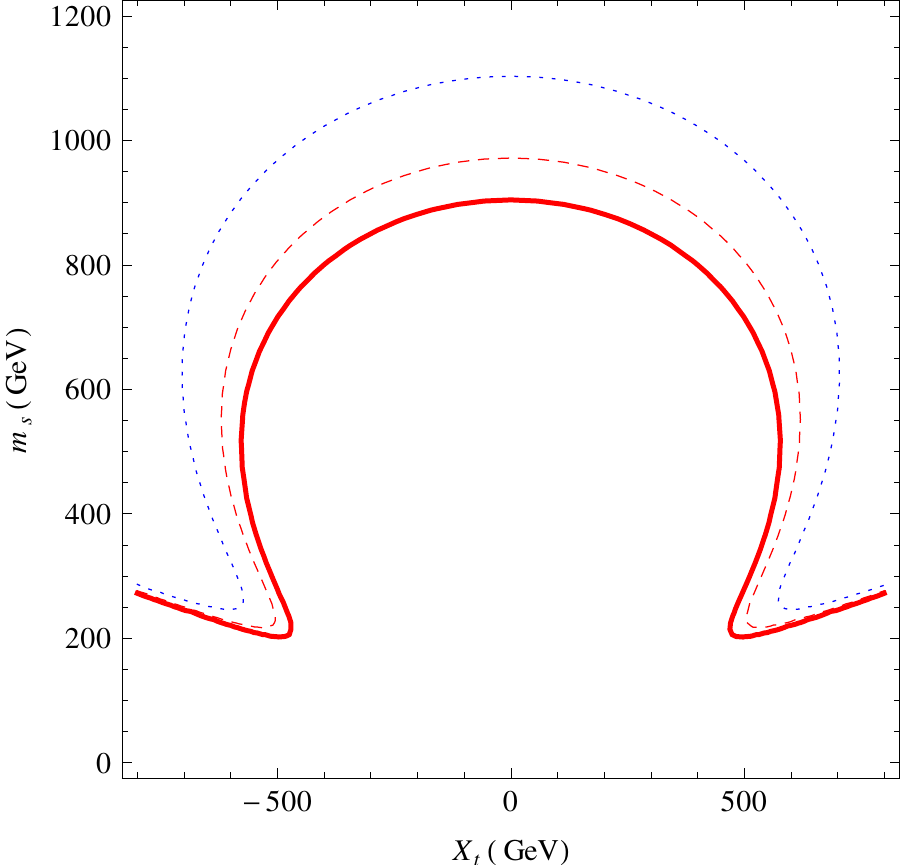}
\caption{Contour plot   of $m_h=125$GeV in the $X_t \,\mbox {versus} \, m_s$
plane with $\lambda_s =-0.21$, $Y_\nu=0.21$ (blue dotted $tan (\beta) =5$), (red
dashed  $tan (\beta)=7$)  and (red solid $tan (\beta) =10$).}
 \label{fig2}
\end{center}
\end{figure}
\begin{figure}[tb]
\begin{center}
	\includegraphics[scale=0.96]{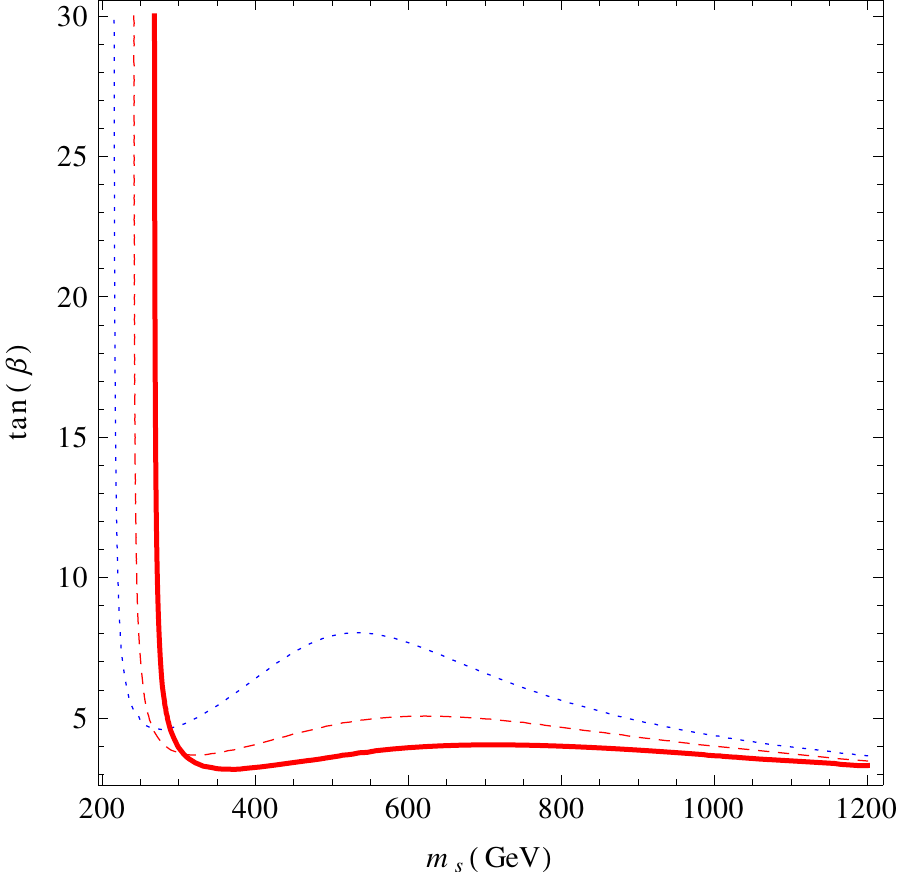}
\caption{Contour plot of $m_h=125$ GeV  in the $tan (\beta) \, \mbox {versus} \,
m_s$ plane  with $\lambda_s =-0.21$, $Y_\nu=0.21$ (blue dotted $X_t=600$GeV),
(red dashed $X_t=700$GeV) and (red solid $X_t=800$GeV). }
 \label{fig3}
\end{center}
\end{figure}
\end{document}